# Brownian motion of charged particles driven by correlated noise

V. Lisy and J. Tothova

Department of Physics, Technical University of Kosice, Park Komenskeho 2,
04200 Kosice, Slovakia

Electronic addresses: vladimir.lisy@tuke.sk, jana.tothova@tuke.sk

**Abstract**

Stochastic motion of charged particles in the magnetic field was first studied almost half a century ago in the classical works by Taylor and Kurşunoğlu in connection with the diffusion of electrons and ions in plasma. In their works the long-time limits of the mean square displacement (MSD) of the particles have been found. Later Furuse on the basis of standard Langevin theory generalized their results for arbitrary times. The currently observed revival of these problems is mainly related to memory effects in the diffusion of particles, which appear when colored random forces act on the particles from their surroundings. In the present work an exact analytical solution of the generalized Langevin equation has been found for the motion of the particle in an external magnetic field when the random force is exponentially correlated in the time. The obtained MSD of the particle motion across the field contains a term proportional to the time, a constant term, and contributions exponentially decaying in the time. The results are more general than the previous results from the literature and are obtained in a considerably simpler way applicable to many other problems of the Brownian motion with memory.

## I. INTRODUCTION

The stochastic motion of charged particles in external magnetic fields is an old problem, the interest to which appeared in connection with the diffusive processes in plasma. In the work [1] Taylor considered the diffusion of ions in plasma across the magnetic field with the stochasticity arising from the fluctuations of the electric field. The equations of motion for the ions were the Langevin equations with the dynamic friction on an ion proportional to its velocity. The solution for the mean square displacement (MSD) was found in the limit of long times. Kurşunoğlu [2, 3] derived the formulas for the diffusion of charged particles across and along the magnetic field as for the motion of Brownian particles with fluctuating accelerations and found the MSD from the probability density distribution for the particle velocity. In the paper [3] also a possible anisotropy in the dynamical friction has been considered by introducing the friction anisotropy matrix. In this case the MSD, obtained also for long times, and the diffusion coefficients in two perpendicular directions across the field could be different. Kurşunoğlu used a useful method to evaluate the mean square velocity and the MSD at equilibrium, however, a complete characterization of the stochastic process was not given. This was done by Czopnik and Garbazcewski in Ref. [4], where the transition probability densities governing the velocity, phase-space and the configuration space processes have been derived. For a further discussion see the works [4 - 7]. The generalization of the theory [1 - 3] for arbitrary time intervals has been done by Furuse [8]. Following the Ornstein-Uhlenbeck and the Fokker-Planck method, Furuse calculated the probability distribution for the Brownian particle coordinates and velocities. In the diffusion process of long times the limiting equations agree with the results by Kurşunoğlu [2, 3] for the diffusion of ions across the stationary magnetic field. It should be noted that in all the discussed papers the friction force that the particle experiences during its motion is the Stokes force and the stochastic force in the Langevin equation is the white-noise (delta-correlated) force. However, such a description of the Brownian motion in fluids is appropriate only for steady motion, *i.e*., at long times. Alternatively, as first pointed out by Vladimirsky and Terletzky [11], it is valid for short times but for particles with the density much larger than that of the surrounding fluid. In a more general description the friction force should reflect the memory in the particle dynamics. Accordingly, the fluctuation-dissipation theorem dictates that the stochastic force in the Langevin equation must be changed to a colored-noise force [12]. This results in a generalized Langevin equation, which becomes a Volterra-type integro-differential stochastic equation. The solution of this equation for a charged Brownian particle moving in a fluid in the magnetic field can be found in the work by Karmeshu [13] (for the corrections of this solution see [14, 15]). In this case the so called hydrodynamic memory or viscous aftereffect (see [15] and references therein) plays a role. It naturally arises within the nonstationary Navier-Stokes hydrodynamics and for particles moving in a liquid it reveals in the appearance of a resistance force (the Boussinesq-Basset force) that in the time *t* depends on the state of the particle motion in all the preceding moments of time. A different kind of memory, the case when the memory kernel in the generalized Langevin equation exponentially decreases in time, was studied in Ref. [6]. Among other papers that represent the currently observed renewal of the interest to the problems of the dynamics of charged Brownian particles in external fields the works [17, 18] should be mentioned. In Ref. [17], the motion of a charged Brownian particle in a harmonic trap was theoretically described in the case when the trap is dragged with a constant velocity and when the particle is subjected to an ac force. The work distribution of the particle has been solved analytically and the Jarzynski equality [19] was verified. In addition, it was shown that this equality is consistent with the Bohr - van Leeuwen theorem on the absence of diamagnetism in a classical system. In a recent paper [18] (containing a number of relevant citations on the theme), a similar problem was considered for a trapped Brownian particle under the action of a constant magnetic field and a time-dependent electric field. The main aim of that work was to prove the validity of the stationary state fluctuation theorem [20] and the transient fluctuation theorem [21]. Note that in Refs. [17, 18] the equations of motion for the particle were chosen in the form of the usual Langevin equations with the Stokes friction. It would be thus of interest to develop these works to the systems with memory.

The aim of the present paper is to give a solution of the generalized Langevin equation for the charged Brownian particle in the constant external magnetic field in the case when the random force driving the particle is a colored noise force. This force is chosen in the form that corresponds to the Ornstein-Uhlenbeck process, *i.e.*, its time correlation function exponentially decreases in time. Our results of Section III correct the work [6] where the same problem was considered for the first time. Before these calculations, in Section II we obtain the classical results by Furuse [8], coming from the usual Langevin equation. The reason why we recalculate these results is the demonstration of a method that allowed us to reach the solution in a very simple way. For the motion along the field this method is based on the rule formulated long ago by Vladimirsky [22] that allows rewriting the Langevin equation for the position of the particle to the equation for its MSD. The motion across the field is described by two coupled equations that we convert to equations for the velocity autocorrelation functions in the plane perpendicular to the field. In linear approximation, our approach is applicable to the description of the Brownian motion with any kind of memory.

## II. USUAL LANGEVIN EQUATION FOR THE BROWNIAN MOTION IN MAGNETIC FIELD

Let us consider the motion of a charged particle in the constant external magnetic field. We begin with the simplest case when the resistance force against the particle motion is modeled as a Stokes friction force - $F(t)$. If $Q$ is the charge of the particle of mass $M$ and $B$ is the constant induction of magnetic field along the axis *z*, then the well known deterministic equation for the particle will be

$$M\dot{\boldsymbol{v}}(t) + \boldsymbol{F}(t) = Q\boldsymbol{v} \times \boldsymbol{B}. \tag{1}$$

For a spherical particle $\boldsymbol{F}(t)$ takes the form $\boldsymbol{F}(t) = \gamma \boldsymbol{v}$ with the constant friction coefficient $\gamma = 6\pi\eta R$, where $R$ is the particle radius and $\eta$ is the viscosity of the solvent. For the phenomenological description of the chaotic motion of mesoscale Brownian particles, in Eq. (1) a random Langevin force should be added. As mentioned in Introduction, for arbitrary time intervals *t* the resulting Langevin equation was first solved by Furuse [8]. As usually, he used the white-noise force $f(t)$ with zero mean and the correlation properties $\langle f_i(t) f_j(t')\rangle = 2k_B T \gamma \delta_{ij} \delta(t - t')$, where $k_B$ is the Boltzmann's constant. Below we give the solution of Eq. (1) without the explicit use of these properties.

Let us denote the MSD of the particle in the *x* direction as $\xi_x(t) = \langle \Delta x^2(t) \rangle = \langle [x(t) - x(0)]^2 \rangle$. Analogously, $\xi_y$ and $\xi_z$ are for the MSD in the directions *y* and *z*. The projection of Eq. (1) onto the axis *z* does not contain the magnetic force so that along the field we have the motion of a free Brownian particle described by the equation

$$M\dot{v}_z = -\gamma v_z + f_z. \tag{2}$$

Its solution can be obtained in a standard way and is familiar from many textbooks. An easy way to obtain it is as follows. Equation (2) can be rewritten for the quantity $V_z$ defined as $V_z = d\xi_z/dt$ in the form of the non-stochastic differential equation

$$M\dot{V}_z = -\gamma V_z + 2k_B T, \tag{3}$$

which has to be solved with the condition $V_z(0) = 0$ (this condition follows also from the definition of the MSD). This method that can be called the Langevin-Vladimirsky rule was established in Ref. [22] for linear Langevin equations not only of the type (2), but for any kind of memory in the system. Here, Eq. (3) is easily obtained if (2) is multiplied by $v_z(0)$ and then statistically averaged (for more details see below after Eq. (6)). One immediately obtains from Eq. (3)

$$V_z(t) = \frac{2k_B T}{\gamma}\left[1 - \exp\left(-\frac{t}{\tau_M}\right)\right], \tag{4}$$

where $\tau_M = M/\gamma = 1/\omega_M$ is the Brownian relaxation time for particles of the mass *M*. The time-dependent diffusion coefficient is $D_z(t) = V_z(t)/2$ and converges to $D_z = k_B T/6\pi\eta R$ as *t* $\to \infty$. The velocity autocorrelation function (VAF) is $\phi_z(t) = \dot{D}_z(t) = (k_BT/M)\exp(-t/\tau_M)$ and the MSD along the field, $\xi_z(t) = \int_0^t V_z(t')dt'$, is

$$\xi_z(t) = \frac{2k_BT}{\gamma}\left\{t - \tau_M\left[1 - \exp\left(-\frac{t}{\tau_M}\right)\right]\right\}. \tag{5}$$

The motion across the field can be considered by the following way, which is much simpler than the methods presented in the literature so far (*e.g.* [4, 8, 23]).

Two equations for the motion across the field are

$$\dot{\upsilon}_x = \omega_c \upsilon_y - \omega_M \upsilon_x + \frac{f_x}{M}, \quad \dot{\upsilon}_y = -\omega_c \upsilon_x - \omega_M \upsilon_y + \frac{f_y}{M}, \tag{6}$$

where $\omega_c = QB/M = 1/\tau_c$ is the cyclotron frequency. We multiply both these equations by $\upsilon_x(0)$ and statistically average. Since the random initial value $\upsilon_x(0)$ and the values of the forces $f_x$ and $f_y$ at the time $t$ are statistically independent, we obtain

$$\dot{\phi}_x(t) = -\omega_M \phi_x(t) + \omega_c \langle \upsilon_y(t)\upsilon_x(0)\rangle, \tag{7}$$

$$\frac{d}{dt}\langle \upsilon_y(t)\upsilon_x(0)\rangle = -\omega_M \langle \upsilon_y(t)\upsilon_x(0)\rangle - \omega_c \phi_x(t).$$

The autocorrelation function for the *x*-component of the velocity, $\phi_x(t) = \langle \upsilon_x(t)\upsilon_x(0)\rangle$, thus obeys the equation

$$\ddot{\phi}_x + 2\omega_M \dot{\phi}_x + (\omega_c^2 + \omega_M^2)\phi_x = 0. \tag{8}$$

Analogously, the same equation is obtained for $\phi_y(t) = \langle \upsilon_y(t)\upsilon_y(0)\rangle$. The initial conditions follow from the equipartition theorem, $\phi_i(0) = k_BT/M$, and Eq. (7), $\dot{\phi}_i(0) = -\omega_M k_BT/M$, $i = x, y$, so that the solution for the VAF in $x$ and $y$ directions is

$$\phi_i(t) = \frac{k_BT}{M}\exp(-\omega_M t)\cos(\omega_c t). \tag{9}$$

Having the VAF, other relevant time correlation functions for the Brownian particle can be easily found. The time-dependent diffusion coefficient in the $x$ direction, $D_x(t)$, is connected to the VAF by the relation $\phi_x(t) = dD_x(t)/dt$, and the MSD $\xi_x(t) = \langle[x(t) - x(0)]^2\rangle$ is obtained as [24]

$$\xi_x(t) = 2\int_0^t (t-s)\phi_x(s)ds, \tag{10}$$

so that $\phi_x(t) = \ddot{\xi}_x(t)/2$. Equations (4) and (5) are easily derived representing the distance a particle moves in time as an integral of its velocity $\upsilon_x(t)$,

$$x(t) - x(0) = \int_0^t \upsilon_x(s)ds. \tag{11}$$

The full VAF, since no magnetic rotational effect is present in the Brownian motion in the plane perpendicular to the field, is $\phi_{xy}(t) = \phi_x(t) + \phi_y(t) = 2\phi_x(t)$. The time-dependent diffusion coefficient

$$D_{xy}(t) = \frac{2k_B T}{M} \frac{1}{\omega_M^2 + \omega_c^2}\left[\omega_M + \exp(-\omega_M t)(\omega_c \sin \omega_c t - \omega_M \cos \omega_c t)\right] \tag{12}$$

and the MSD

$$\xi_{xy}(t) = \frac{4k_B T}{M} \frac{1}{\omega_M^2 + \omega_c^2}\left\{\omega_M t + \frac{\omega_c^2 - \omega_M^2}{\omega_M^2 + \omega_c^2}\left[1 - \exp(-\omega_M t)\left(\frac{2\omega_M \omega_c}{\omega_c^2 - \omega_M^2}\sin \omega_c t + \cos \omega_c t\right)\right]\right\}. \tag{13}$$

agree with the results [4, 8, 23], where they were obtained in a much more involved way.

## III. GENERALIZED LANGEVIN EQUATION WITH THE EXPONENTIALLY CORRELATED NOISE

The simple method presented in the previous section where the known final results have been obtained for memoryless motion is applicable to more interesting situations of the Brownian motion with memory, allowing to find new results. Let us now consider such a case when the memory effect during the particle motion is taken into account using a generalized Langevin equation. For the charged Brownian particle in the magnetic field the equation of motion (1) now changes to

$$M\dot{\upsilon}(t) + \int_0^t \Gamma(t-t')\upsilon(t')dt' = Q\upsilon \times \boldsymbol{B} + \boldsymbol{\zeta}(t). \tag{14}$$

Here we will study the situation when the memory kernel $\Gamma$ has the Ornstein-Uhlenbeck form $\Gamma(t) = (\gamma^2/m)\exp(-\gamma t/m) = (\gamma/\tau_m)\exp(-t/\tau_m)$, $\tau_m = m/\gamma$. Then, in accordance with the fluctuation-dissipation theorem [12, 25 - 27], the stochastic force $\boldsymbol{\zeta}(t)$ corresponds to $m\dot{\boldsymbol{u}}(t)$, where $\boldsymbol{u}(t)$ is the solution of the usual Langevin equation $m\dot{\boldsymbol{u}}(t) + \gamma \boldsymbol{u}(t) = \boldsymbol{f}(t)$ with the random force having the same statistical properties as in the previous section. That is, $\langle \zeta_i(t)\zeta_j(t')\rangle = \delta_{ij}k_B T\Gamma(t-t')$.

Equation (14) for the motion along the field does not contain the magnetic field and can be as in the previous section directly converted to the Vladimirsky equation

$$M\dot{V}_z + \frac{\gamma}{\tau_m}\int_0^t \exp\left[-(t-t')/\tau_m\right]V_z(t')dt' = 2k_BT, \tag{15}$$

with the same initial condition $V_z(0) = 0$ as in the previous section. By taking the Laplace transform we obtain for $\tilde{V}_z(s) = \mathcal{L}\{V_z(t)\}$

$$\tilde{V}_z(s) = \frac{2k_BT}{M}\frac{1}{\alpha_2-\alpha_1}\left(1+\frac{1}{\tau_m s}\right)\left(\frac{1}{s-\alpha_2}-\frac{1}{s-\alpha_1}\right), \tag{16}$$

where $\alpha_{1,2} = (\gamma/2m)(1\mp\sqrt{1-4m/M})$ are the roots of the equation $s^2 + s/\tau_m + \gamma/(M\tau_m) = 0$. The inverse transform yields [28]

$$V_z(t) = \frac{2k_BT}{M}\left\{\frac{1}{\alpha_1\alpha_2\tau_m} + \frac{1}{\alpha_2-\alpha_1}\left[\left(1+\frac{1}{\alpha_2\tau_m}\right)\exp(\alpha_2 t) - \left(1+\frac{1}{\alpha_1\tau_m}\right)\exp(\alpha_1 t)\right]\right\}. \tag{17}$$

The MSD is obtained by simple integrating of this equation, $\xi_z(t) = \int_0^t V_z(t')dt'$. Using $\alpha_1\alpha_2 = \gamma^2/(mM)$, it is seen that the first term in {} is the Einstein long-time limit $2k_BT/\gamma$. The time-dependent diffusion coefficient is $D_z(t) = V_z(t)/2$. At short times we have the expected result, which is independent of a concrete form of $\Gamma(t)$,

$$D_z(t) \approx \frac{k_BT}{M}t, \quad \xi_z(t) \approx \frac{k_BT}{M}t^2, \quad t \to 0. \tag{18}$$

If we denote $\mu = \sqrt{1-4m/M}$, $D_z(t)$ can be given a compact form

$$D_z(t) = \frac{k_BT}{\gamma}\left\{1 - \frac{1}{4\mu}\exp(-\gamma t/2m)\left[(1+\mu)^2\exp(\gamma\mu t/2m) - (1-\mu)^2\exp(-\gamma\mu t/2m)\right]\right\}. \tag{19}$$

In the special case when $M < 4m$, i.e. when the roots $\alpha$ are complex, the solution describes damped oscillations. In the overdamped limit (if $M << 4m$), we have for $D_z(t)$

$$D_z(t) \approx \frac{k_BT}{\gamma}\left\{1 + \exp(-\gamma t/2m)\left[\sqrt{\frac{m}{M}}\sin\left(\frac{\gamma t}{\sqrt{mM}}\right) - \cos\left(\frac{\gamma t}{\sqrt{mM}}\right)\right]\right\}. \tag{20}$$

Now, consider the motion across the field. The equations that follow from Eq. (14) for the $x$ and $y$ components of the velocity differ only by the sign of the term proportional to $Q$:

$$M\frac{dv_x}{dt} = QBv_y - \int_0^t \Gamma(t-t')v_x(t')dt' + \zeta_x(t), \tag{21}$$

$$M\frac{dv_y}{dt} = -QBv_x - \int_0^t \Gamma(t-t')v_y(t')dt' + \zeta_y(t).$$

After multiplying both these equations by $v_x(0)$ and ensemble averaging we obtain two equations for the VAF $\phi_x(t)$

$$\dot{\phi}_x(t) = \frac{1}{\tau_c}\langle v_y(t)v_x(0)\rangle - \frac{1}{\tau_m\tau}\int_0^t dt'\phi_x(t')\exp[-(t-t')/\tau_m], \tag{22}$$

$$\frac{d}{dt}\langle v_y(t)v_x(0)\rangle = -\frac{1}{\tau_c}\phi_x(t) - \frac{1}{\tau_m\tau}\int_0^t dt'\langle v_y(t')v_x(0)\rangle\exp[-(t-t')/\tau_m],$$

with $\tau = M/\gamma$. Using the Laplace transformation and the initial conditions $\phi_x(0) = k_B T/M$ and $\langle v_x(0)v_y(0)\rangle = 0$, we obtain for the quantities $\tilde{\phi}_x(s) = \mathcal{L}\{\phi_x(t)\}$ and $\tilde{v}_{xy}(s) = \mathcal{L}\{\langle v_x(0)v_y(t)\rangle\}$

$$s\tilde{\phi}_x(s) - \frac{k_B T}{M} = \frac{\tilde{v}_{xy}(s)}{\tau_c} - \frac{1}{\tau_m\tau}\frac{\tilde{\phi}_x(s)}{s+\tau_m^{-1}}, \quad s\tilde{v}_{xy}(s) = -\frac{1}{\tau_c}\tilde{\phi}_x(s) - \frac{1}{\tau_m\tau}\frac{\tilde{v}_{xy}(s)}{s+\tau_m^{-1}}. \tag{23}$$

Analogous equations are obtained from (21) for the quantities $\tilde{\phi}_y(s)$ and $\tilde{v}_{yx}(s) = \mathcal{L}\{\langle v_y(0)v_x(t)\rangle\}$. These sets of equations have the same solutions for $i = x, y$:

$$\tilde{\phi}_i(s) = \frac{k_B T}{M}\left(s + \frac{1}{\tau_m}\right)\frac{\psi(s)}{\psi^2(s) + \tau_c^{-2}(s+\tau_m^{-1})^2}, \tag{24}$$

where $\psi(s) = s^2 + s/\tau_m + 1/(\tau_m\tau)$. For the time-dependent diffusion coefficient in the plane perpendicular to the magnetic field, $D_{xy}(t) = D_x(t) + D_y(t) = 2\int_0^t dt'\phi_x(t')$, we thus have $\tilde{D}_{xy}(s) = 2s^{-1}\tilde{\phi}_x(s)$. The inverse Laplace transformation of this quantity is obtained after the decomposition in simple fractions,

$$\tilde{D}_{xy}(s) = \frac{k_B T}{M}\left(1+\frac{1}{\tau_m s}\right)\left\{\frac{1}{\beta_2-\beta_1}\left(\frac{1}{s-\beta_2}-\frac{1}{s-\beta_1}\right) + \frac{1}{\beta_4-\beta_3}\left(\frac{1}{s-\beta_4}-\frac{1}{s-\beta_3}\right)\right\}, \tag{25}$$

where $\beta_{1,2}$ and $\beta_{3,4}$ are the roots of the equations $\psi(s) + i\tau_c^{-1}(s+\tau_m^{-1}) = 0$ and $\psi(s) - i\tau_c^{-1}(s+\tau_m^{-1}) = 0$, respectively. We find

$$2\beta_{1,2} = -\left(\frac{1}{\tau_m}+\frac{i}{\tau_c}\right) \pm \sqrt{\left(\frac{1}{\tau_m}-\frac{i}{\tau_c}\right)^2 - \frac{4}{\tau_m\tau}}, \tag{26}$$

and $\beta_{3,4}$ are given by an analogous formula with $-\tau_c$ (or $\beta_{3,4} = \beta_{1,2}^*$). The result in the $t$-representation is expressed through the exponential functions $\exp(\beta_i t)$ that converge to zero as $t \to \infty$. In the long-time limit $D_{xy}(t)$ and the MSD across the field, $\xi_{xy}(t) = 2\int_0^t D_{xy}(t')dt'$, thus do not depend on $m$,

$$D_{xy}(t) \approx \frac{k_B T}{M\tau_m}\left(\frac{1}{\beta_1\beta_2} + \frac{1}{\beta_3\beta_4}\right) = \frac{2k_B T}{\gamma}\frac{1}{1+(\tau/\tau_c)^2}, \quad \xi_{xy}(t) \approx \frac{4k_B T}{\gamma}\frac{t}{1+(\tau/\tau_c)^2}. \tag{27}$$

At the $B = 0$ limit, $\xi_{xy}(t) \approx 4k_B T t/\gamma$. Note that the correct one-dimensional $D_i(t \to \infty)$ is two times smaller than in Ref. [6] (Eq. (19)). The result for all times can be obtained analytically in a very simple way. First, the inverse Laplace transformation of Eq. (25) gives [28]

$$D_{xy}(t) = \frac{2k_B T}{M}\text{Re}\left\{\frac{1}{\beta_1\beta_2\tau_m} + \frac{1}{\beta_2 - \beta_1}\left[\left(1 + \frac{1}{\beta_2\tau_m}\right)\exp(\beta_2 t) - \left(1 + \frac{1}{\beta_1\tau_m}\right)\exp(\beta_1 t)\right]\right\}. \tag{28}$$

In spite of the fact that the constants $\beta_i$ from Eq. (26) contain square root of complex argument and are thus two-valued, the final results for the correlation functions are one-valued. It is assured by the structure of the results and can be checked by representing the root in $\beta_{1,2}$ in the form $\pm r^{1/2}\exp(i\varphi/2)$. The choice of the sign + gives $2\beta_1^+ = -1/\tau_m - i/\tau_c + r^{1/2}\exp(i\varphi/2)$ and $2\beta_2^+ = -1/\tau_m - i/\tau_c - r^{1/2}\exp(i\varphi/2)$. For the opposite sign we have $\beta_1^- = \beta_2^+$ and $\beta_2^- = \beta_1^+$. Both the choices thus lead to the same $D_{xy}(t)$ in Eq. (28). Integrating this equation we obtain the MSD

$$\xi_{xy}(t) = \frac{4k_B T}{M}\text{Re}\frac{1}{\beta_1\beta_2}\left\{\frac{t}{\tau_m} + \frac{1}{\beta_2 - \beta_1}\left[\beta_1\left(1 + \frac{1}{\beta_2\tau_m}\right)(e^{\beta_2 t} - 1) - \beta_2\left(1 + \frac{1}{\beta_1\tau_m}\right)(e^{\beta_1 t} - 1)\right]\right\}. \tag{29}$$

Thus, $\xi_{xy}(t)$ contains, next to the 'Einstein' term proportional to $t$, a constant term $4k_B T\gamma^2(M-m)[(\tau/\tau_c)^2 - 1][(\tau/\tau_c)^2 + 1]^{-2}$. After dividing by $(4k_B T\tau/\gamma)[1+(\tau/\tau_c)^2]^{-1}$, the normalized MSD from (29), $\bar{\xi}_{xy}(t)$, depends on the dimensionless time $t/\tau$ and contains only two parameters $\tau/\tau_c$ and $\tau/\tau_m$ that characterize the studied system (one more parameter $\tau_m/\tau_c$ is expressed through these two constants). It is easy to check that in the absence of the magnetic field the constants $\beta_{1,2}$ reduce to $\alpha_{1,2}$ from Eq. (16). Equation (28) thus becomes identical with (17) and Eq. (29) will be $\xi_{xy}(t) = 2\xi_z(t)$, where $\xi_z(t)$ is for the MSD along the field.

The classical results for particles driven by the white noise are obtained in the limit of zero correlation time of the colored noise. When $\tau_m \to 0$, the constants $\beta_{1,2}$ become $\beta_1 \approx -(1/\tau + i/\tau_c)$ and $\beta_2 \approx -1/\tau_m + 1/\tau$, so that the exponential function $\exp(\beta_2 t)$ tends to zero (if $t$

> 0). The resulting VAF, $\phi_x(t) = \phi_y(t) = (k_B T / M)\exp(-t/\tau)\cos(t/\tau_c)$, valid for all $t \geq 0$, is exactly the one that follows from the Langevin equation with the white noise force. Consequently, Eqs. (28) and (29) correspond to (12) and (13), respectively, and when $B = 0$ they agree with Eq. (5) along the magnetic field. This correspondence can be obtained also directly from Eqs. (28) and (29), using $\beta_1 \beta_2 \tau_m \to \tau^{-1} + i\tau_c^{-1}$ and $(\beta_2 - \beta_1)^{-1} \approx -\tau_m$, as $\tau_m \to 0$. Then, e.g. for $D_{xy}$, we return to Eq. (12),

$$D_{xy}(t) \approx \frac{2k_B T}{M} \mathrm{Re}\left\{\left(\frac{1}{\tau} + \frac{i}{\tau_c}\right)^{-1}\left[1 - \exp\left(-\frac{t}{\tau} - \frac{it}{\tau_c}\right)\right]\right\}$$

$$= \frac{2k_B T}{M} \frac{1}{\tau^{-2} + \tau_c^{-2}}\left\{\frac{1}{\tau} - \exp\left(-\frac{t}{\tau}\right)\left(\frac{1}{\tau}\cos\frac{t}{\tau_c} + \frac{1}{\tau_c}\sin\frac{t}{\tau_c}\right)\right\}. \quad (30)$$

At long times the exponential functions are negligible and we have from Eq. (29)

$$\bar{\xi}_{xy}(t) \approx \frac{t}{\tau} + \left(1 - \frac{\tau_m}{\tau}\right)\frac{(\tau/\tau_c)^2 - 1}{(\tau/\tau_c)^2 + 1} + \ldots \quad (31)$$

As $t \to 0$, the independence on $B$ and the ballistic behavior of the MSD are expected. This really follows from Eq. (29): $\xi_{xy} \approx (2k_B T / M)t^2$. The time-dependent diffusion coefficient $D_{xy}(t \to 0)$ is proportional to $t$, which again disagrees with the result obtained from Eq. (18) in Ref. [6]: if one uses Eqs. (10) and (13) [6] for $g(t)$ and $G(t)$, respectively, and substitutes their short-time limits $g(t) \approx 1 + i\omega t - (t^2/2)(\omega^2 + \gamma'/\tau)$ and $G(t) \approx t + i\omega t^2/2$ in Eq. (18), up to the terms $\sim t^2$ the result is zero (we denote as $\gamma'$ the quantity that is connected to our $\gamma$ by the relation $\gamma = \gamma' M$).

Figures 1 – 5 show the numerical calculations of the normalized MSD $\bar{\xi}_{xy}(t)$ as a function of $t/\tau$ using Mathematica [29]. We used the parameters $\tau_c = \tau_m$ except Fig. 5 calculated for the realistic case of a micrometer-sized strongly charged colloidal particle with the surface charge density $\sigma$ of some tenths of C/m$^2$ [30]. Such particles with the radius $R = 1$ μm and $\sigma = 0.5$ C/m$^2$ carry the charge $Q = 6\cdot10^{-12}$ C. We assume that the external magnetic field is $B = 1$ T and the driving particles in water have a radius $r = 0.1$ μm. The friction factor $\gamma = 10^{-9}$ kg·s$^{-1}$ is calculated from the Stokes-Einstein formula $\gamma = 6\pi\eta r$, where $\eta$ is the viscosity of water. The masses of the particles relates as $M/m = 10^3$.

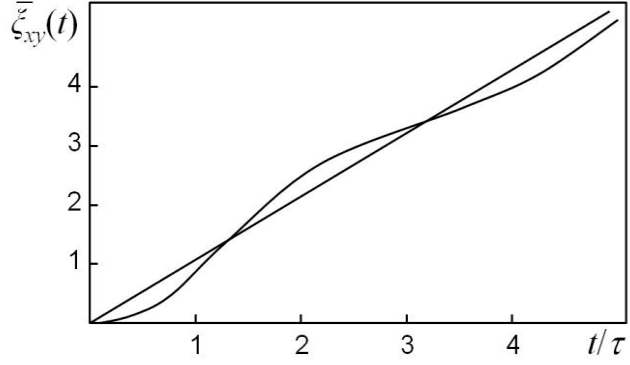

**Fig. 1.** Normalized MSD as a function of time, calculated from Eqs. (29) and (26) with the parameters $\tau/\tau_c = QB/\gamma = \tau/\tau_m = M/m = 1$. The Einstein limit (straight line) is shown for comparison.

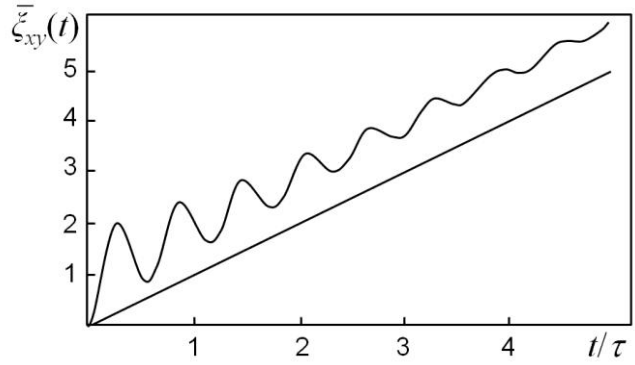

**Fig. 2.** The same as in Fig. 1 with $\tau/\tau_c = \tau/\tau_m = 10$.

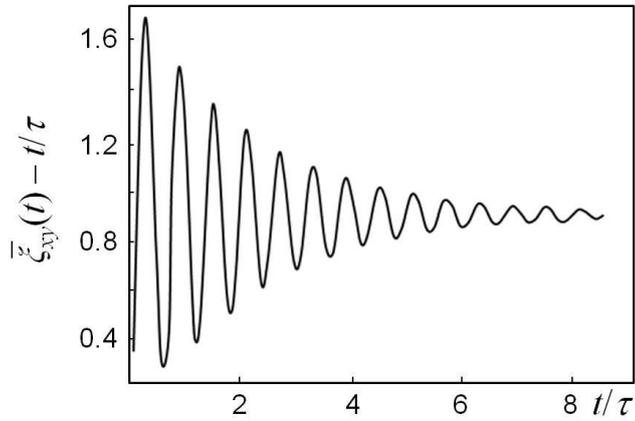

**Fig. 3.** Normalized MSD from Fig. 2 minus the Einstein term $t/\tau$. At long times the graph corresponds to Eq. (31): $\bar{\xi}_{xy}(t) - t/\tau$ converges to a constant.

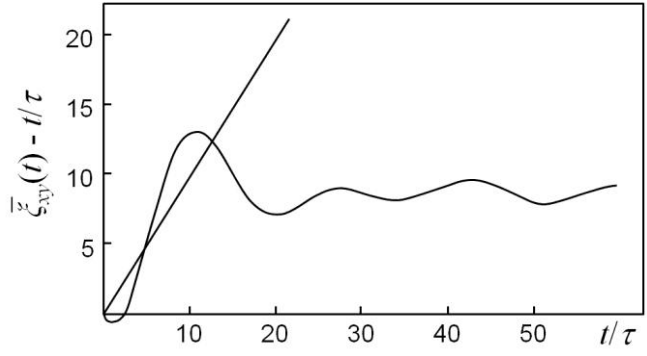

**Fig. 4.** The same as in Fig. 1 with $\tau/\tau_c = \tau/\tau_m = 0.1$.

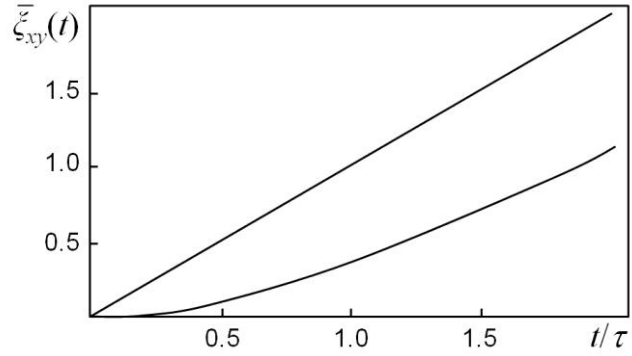

**Fig. 5.** The same as in Fig. 1 with $\tau/\tau_c = 3\cdot10^{-3}$ and $\tau/\tau_m = 10^3$. This choice corresponds to realistic charged colloidal particles as described in the text.

## IV. CONCLUSION

In recent years, a number of papers appeared, in which the classical works by Taylor and Kurşunoğlu on the motion of charged particles in a fluctuating environment and in a magnetic field have been developed. One of the interesting directions in these studies concerns the account for the memory effects in the particle motion, which was also the main aim of the present paper. Basically, there are two approaches to the solution of such problems - coming from the Langevin equation for the dynamic variables of the particle [12], or solving the associated Fokker-Planck (or Kramers) equation for their probability density [31]. Here we oriented on the solution of the Langevin equation. This was done using a physically clear method in the spirit of the original work by Langevin [32] that allowed us to reach the solution in a significantly simpler way than in the previous works in the literature. The method is applicable to linear Langevin equations in their standard form (when the dissipative term in the equation of motion is the Stokes friction force), as well as to the generalized Langevin equations (when the friction is modeled by an integral representing the convolution of the memory kernel and the particle velocity). Both these cases were exactly solved in this work; in the latter case with a particular memory kernel exponentially decreasing in time. The obtained mean square displacement of the particle motion across the field contains an 'Einstein' term proportional to the time, a constant term, and exponentially decaying contributions. These results correct the previous work [6]. In the limits of short and long times the mean square displacement is in the main approximation independent of the correlation time of the driving stochastic force. At short times there is also no dependence on the magnetic field and the motion is ballistic. It is shown for arbitrary times that no magnetic rotational effect is present in the Brownian motion across the field.

## ACKNOWLEDGMENTS

This work was supported by the Agency for the Structural Funds of the EU within the projects NFP 26220120021 and 26220120033, and by the grant VEGA 1/0300/09.